\documentclass[a4paper,fleqn]{cas-dc}
\usepackage[sort&compress,numbers]{natbib}
\usepackage{amsthm}
\usepackage{setspace}
\usepackage{caption}
\usepackage{subcaption}

\newcommand{\mcal}[1]{\mathcal{#1}}

\begin{document}
\let\WriteBookmarks\relax
\def\floatpagepagefraction{1}
\def\textpagefraction{.001}
\shorttitle{Global Profits, Local Decisions: Why Global Cooperation Falters in Multi-level Games}
\shortauthors{J. Zhao et~al.}

\title [mode = title]{Global Profits, Local Decisions: Why Global Cooperation Falters in Multi-level Games}        

\author[1]{Jinhua Zhao}[orcid=0000-0003-4290-6540]
\cormark[1]
\ead{jzhao1101@ccnu.edu.cn}
\address[1]{Central China Normal University Wollongong Joint Institute, Faculty of Artificial Intelligence in Education, Central China Normal University, Wuhan 430079, China}
\address[2]{School of Economics and Management, Wuhan University, Wuhan 430072, China}

\author[1]{Xinguo Yu}[orcid=0000-0001-8379-6742]
\ead{xgyu@ccnu.edu.cn}

\author[2]{Rui Ding}[orcid=0000-0001-6248-001X]
\ead{dingrui@whu.edu.cn}

\author[3]{Cuiling Gu}[orcid=0000-0002-8575-9889]
\ead{gucui036@haut.edu.cn}

\author[2]{Xianjia Wang}
\ead{wangxj@whu.edu.cn}

\address[3]{School of Management, Henan University of Technology, Henan 450001, China}
\cortext[cor1]{Corresponding author}

\begin{abstract}
Global cooperation often falters despite shared objectives, as misaligned interests and unequal incentives undermine collective efforts, such as those in international climate change collaborations. To tackle this issue, this paper introduces a multi-level game-theoretic model to analyze the dynamics of complex interactions within hierarchical systems. The model consists of global, local, and pairwise games, and two strategy types—binary and level-based strategies—are explored under varying parameter conditions. Using computational simulations and numerical analysis, we examine how factors across different levels influence player decisions, game dynamics and population phase transitions during the evolutionary process. Our findings reveal that although the increase of profit rates at local and pairwise games enhances cooperation within the population, the global game exerts minimal influence on player decisions and population states under both strategy settings. Particularly, analytical and simulation results show that, under binary strategies, global profit does not influence localized decision-making of players, while under level-based strategies, players cooperating at the global level are eventually outcompeted due to the evolutionary disadvantage even when global profit is substantial.
These insights contribute to a theoretical understanding of cooperation dynamics in multi-level systems and may offer implications for fostering global collaboration on challenges like climate change.
\end{abstract}

\ExplSyntaxOn
\keys_set:nn { stm / mktitle } { nologo }
\ExplSyntaxOff
\maketitle
\section{Introduction}
Understanding the emergence and evolution of cooperative behaviors within a population of inherently self-interested individuals with complex relations has long captivated the interest of researchers across various disciplines \citep{Axelrod1981, Kollock1998, Nowak2006, Friedman2016}. The study of cooperative behaviors within the framework of evolutionary dynamics provides valuable insights into how such behaviors emerge and evolve over time. Evolutionary game theory is a powerful tool that explores the strategies employed by individuals in competitive environments, and sheds light on the conditions that foster cooperation and the mechanisms by which it spreads through a population \citep{Friedman2016, Smith1982}. However, in populations such as social networks or geopolitical networks, interactions often hold a multi-level nature, meaning participants engage in games with different groups of opponents of different sizes.

One quintesential example is the global effort to combat climate change. In this complex scenario, nations must navigate a delicate balance between pursuing their own interests and recognizing the shared responsibility for the health of the planet. The challenge of addressing climate change underscores the need for collaboration among nations at multiple levels, from global agreements, regional collaborations to multilateral or bilateral agreements. On the one hand, we have witnessed the effort of forging a global cooperation against climate change such as Kyoto Protocol \citep{Kyoto1997} and Paris Agreement \citep{Paris2015}, despite their underwhelming outcome \citep{Fawcett2015,Rogelj2016, Victor2017,Kuriyama2018, Hartl2019, Maamoun2019}. 
On regional collaboration, nations often engage in cooperative endeavors tailored to their specific geographical, economic, and political contexts. For instance, regions have rolled out initiatives such as the European Union's European Green Deal and Emissions Trading System \citep{ EU2005, EU2020}, and North American Climate, Clean Energy, and Environment Partnership Action Plan \citep{NA2016}. These agreements allow participating nations to pool resources, share knowledge, and implement coordinated strategies to mitigate the impacts of climate change on a more localized scale.
Nations may also sign bilateral or multilateral agreements, establishing mutually beneficial arrangements to address shared environmental concerns or promote sustainable development initiatives. These types of agreements enable nations to collaborate more intimately, focusing on specific areas of cooperation that align with their respective interests and capabilities, such as U.S.-China Joint Announcement on Climate Change in 2014 \citep{UC2014}, Franco-German treaty on energy transition \citep{FG2019}, and Canada-Mexico Action Plan on environment and climate change \citep{CM2022}. 
Additionally, nations have been proposing various climate policies within their respective jurisdictions to encourage actions to slow down the climate change process, and to mitigate its negative effects \citep{Georgeson2017, Espeland2018}.  

On the other hand, researchers have dedicated efforts to modeling and creating mechanisms that incentivize cooperation in combating climate change, via approaches including theoretical analysis \citep{Santos2011, Carfi2019}, numerical and computational simulations \citep{Gois2019, Zhao2021, Zhao2023, Chica2024}, social experiments \citep{Milinski2008, Milinski2011, Wang2020}, statistical physics \citep{Jusup2022, Vazquez2022}, and even decision-support tools \citep{Molina-Perez2023}. They analyzes the equilibria and evolutionary dynamics of corresponding games across various scenarios, and devise mechanisms aimed at promoting cooperation. Different variations of public goods game, collective risk dilemma and other game models have been used to understand the interactions between nations facing climate change, and mechanisms such as reward, sanction, communication and environmental feedback are studied in terms of their efficiency in promoting cooperation \citep{Carfi2019, Gois2019, Wang2020, Zhao2021}. 

Additionally, as a well-studied topic, social dilemmas in structured populations have been extensively studied to explore game equilibria and dynamics, with a focus on diverse player strategies, endogenous factors, exogenous mechanisms, and innovative modeling approaches \citep{Nowak1992, Sigmund2001, Brandt2006, Santos2006, Santos2008, Tarnita2009, Helbing2010, Perc2012, Chen2014, Szolnoki2016, Allen2017, Nanda2017, DU2019, Szolnoki2021, Szolnoki2022, Wang2024}. 

Nontheless, there have been more recent studies dedicated to address similar problems from a multi-level perspective, which consider the impact of factors such as group selection, cross-level communication, exogenous intervention \cite{Szolnoki2021, Guo2023, Li2024}. Besides, preliminary multi-level game theory applications have emerged in fields such as electricity market and hydrogen distribution as well \citep{Budi2021, Zhao2024}. 

In this paper, we aim to study the impact of factors across different levels on the evolution of cooperation within a proposed multi-level framework, covering pairwise interactions, local collaborations, and global collaborations. Of particular interest is how the same factors at different levels can exert distinct influences on the game state and player strategies. Through computational simulations and numeric analysis, we investigate the impact of game parameters at different levels on the overall population state and cooperation tendencies under various strategy settings. By gaining a deeper understanding of these dynamics, we may advance our theoretical understanding of cooperation in multi-level interactions, as well as inspire practical strategies for fostering cooperation in real-world contexts. 

The rest of the paper is organized as follows. In Section \ref{sec:model}, we propose the multi-level game model. 
Section \ref{sec:res} includes the experimental settings, results and analysis. Finally, Section \ref{sec:conclusion} concludes the research work, discusses the potential practical implications of our findings, and points out the limitations and potential future research directions.

\section{Model}\label{sec:model}
We consider a population of players that may engage in games at different levels. 
Particularly, the population of players is denoted by $N$, and a (population structure) graph $G = (N, E)$ is used to represent the relations among the players, where $E \subseteq N^2$ is the set of edges. For any players $i, j \in N$, the edge $(i,j)\in E$ means that they are adjacent and can engage directly in a two-player game. Moreover, for any subset (or group) of players $V \subseteq N$, 
 its associated game is denoted $$g_V = (V, S^V, \pi^V),$$ where $V$ is the set of player, $S^V$ is the set of strategies (or actions, interchangeably) the players can choose from, and $\pi^V$ is the payoff function. 
For each game $g_V$, the set of players $V$ should induce a connected subgraph of $G$, otherwise the disjointed players would neither participate in the game nor affect the outcome. In other words, if we denote the set of edges that have both extremities in $V$ by $E(V)$, $(V, E(V))$ should be a connected subgraph of $G$ for any game $g_V$. Any set $V$ that satisfies this condition can be a {\em viable} group of players. For similar reasons, we also assume that the graph $G$ is connected, since otherwise it can be evaluated separately for each of its connect components. 

This paper considers three different levels of games (and consequently, three levels of viable groups) that might be played within the population. Firstly, any pair of players $i, j \in N$ that are adjacent can engage in a {\em \textbf{pairwise}} game $g_{\{i,j\}}$. 
 Additionally, if we denote by $N(i)$ the set of players containing the player $i \in N$ and all its neighbors, $N(i)$ is a viable group that admits the {\em \textbf{local}} game $g_{N(i)}$. Finally, the whole population engages in the {\em \textbf{global}} game $g_{N}$. 

The set of viable groups that are allowed to engage in a game $g_V$ for a population $N$ is then denoted by $\mcal{V} \subseteq 2^N$. This paper considers the aforementioned pairwise, local and global games, and the set of viable groups that any player $i \in N$ is in can be denoted as 
\begin{align}
\begin{split}
\mcal{V}_i =& \mcal{V}_i^p \cup \mcal{V}_i^l \cup \mcal{V}_i^g  \\
=& \bigl\{\{i,j\}| j\in N(i)\setminus \{i\}\bigr\} \\
 &\cup \bigl\{N(j)|j\in N(i)\bigr\} \\
 &\cup \bigl\{N\bigr\}, 
\end{split}
\end{align}
where $\mcal{V}_i^p$ and $\mcal{V}_i^l$ denote the sets of viable groups of pairwise games and local games for $i$, respectively, and for consistency $\mcal{V}_i^g$ denotes the set containing the only viable group of global game $N$. It is worth noting that, in some cases the viable groups at the three levels of games might overlap, such as in complete graphs where all nodes are adjacent to each other, any local game would be identical as the global game. Such cases can be considered as playing the multi-level game with an adapted strategy space and modified profit rate.

Initially, any player $i \in N$ has $o_i$ units of endowment, which could be distributed among the games it participates in. Therefore, the strategy, or the contribution, of any player in a specific game (in the most general case) is in $S=[0,o_i]$. 

This paper considers two types of strategy settings while assuming that all players have same amount of endowment, i.e., $o_i = 1$ for any $i \in N$. The first setting is binary strategy, where a player can either cooperate or defect in all games it participate in, which means it either contribute all or none of its endowment. The second setting is level-based strategy, where a player decide independently at the three levels of games, allowing them to cooperate at one level and defect at another. For this case we assume the strategies of a player for games at each level is consistent.  
Note that we divide the endowment of each player into three equal parts for the three levels, and then distribute the endowment equally among games within the same level, i.e., $|\mcal{V}_i^p|o_i^p=|\mcal{V}_i^l|o_i^l=o^g=\frac{1}{3}$ by default.

In this paper, games at all three levels are $n$-player public goods game with $n \geq 2$. For any player $i \in N$, and a viable group $V \in \mcal{V}_i$, its endowment and strategy (or contribution) in the game $g_{V}$ is denoted by $o^V_i$ and $s^V_i$, respectively. Its corresponding payoff is then denoted by 
\begin{align} \label{eq:po}
\pi^V (s^V_i, s^V_{-i}) =o_i^V - s_i^V+\frac{r^V}{|V|} \sum\limits_{j\in V}s_j^V,
\end{align}
where $s^V_{-i}$ denotes the strategy profile of all players except $i$ in $V$, $r^V$ denotes the profit rate of public reserve of game $g_V$, and $|V|$ is the number of players in $V$. As this work focuses on the dynamics induced by different levels of games, we also introduce the notation for the payoff for games at each level as follows.
\begin{align}
\pi^p_i &=\sum\limits_{V\in \mcal{V}^p_i}\pi^V (s^V_i, s^V_{-i})\label{eq:po_p}\\
\pi^l_i &= \sum\limits_{V\in \mcal{V}^l_i}\pi^V (s^V_i, s^V_{-i})\label{eq:po_l}\\
\pi^g_i &=\pi^N (s^N_i, s^N_{-i})\label{eq:po_g}
\end{align}
The total payoff of player $i$ is then denoted as 
\begin{align}
\begin{split}
\pi_i &=  \pi^p_i + \pi^l_i  + \pi^g_i\\
&=  \sum\limits_{V\in \mcal{V}_i}\pi^V (s^V_i, s^V_{-i}).
\end{split}
\end{align}
In fact, by modifying the viable groups of games, this multi-level model can represent a variety of games. For instance, for a typical public goods game setup, it can be described as $\mcal{V}_i = \{ N\}$ and $r^{N}=r$ for any $i \in N$. In other words, the players only participate in the global game with a profit rate of $r$. The payoff of player $i$ is then 
\begin{align}
\begin{split}
\pi_i^{PGG} & =\pi^N (s^N_i, s^N_{-i}) \\
& = 1- s_i^N+\frac{r}{|N|} \sum\limits_{j\in N}s_j^N.
\end{split}
\end{align}
In this equation, the part $1- s_i^N$ represents the reserved endowment of player $i$ in the public goods game; $r \sum\limits_{j\in N}s_j^N$ is the total profit of public reserve; $\frac{r}{|N|} \sum\limits_{j\in N}s_j^N$ is then the profit distributed to each player.

In this paper, as players engage in games at three different levels, we consider a default setting of three different profit rates $r^p, r^l, r^g,$ for the pairwise, local and global games, respectively. A summary of notation for the games that any player $i$ can engage in is displayed in Table \ref{tab:games}.
\begin{table*}
\caption{Notation of games any player $i$ can engage in at each level}\label{tab:games}
\begin{tabular}{ccccc}
\hline
Level &  Set of Viable Groups & Endowment &Profit Rate & Payoff\\\hline
Pairwise & $\mcal{V}_i^p = \bigl\{\{i,j\}| j\in N(i)\setminus \{i\} \bigr\}$ & $o_i^p = \frac{1}{3|\mcal{V}^p_i|}$& $r^p$ & $\pi^p_i$\\
Local & $\mcal{V}_i^l = \bigl\{N(j)|j\in N(i)\bigr\}$ &$o_i^l = \frac{1}{3|\mcal{V}^l_i|}$& $r^l$& $\pi^l_i$\\
Global & $ \mcal{V}_i^g = \bigl\{N\bigr\}$ &$o^g = \frac{1}{3}$& $r^g$& $\pi^g_i$\\
\hline
\end{tabular}
\end{table*}

After each round of game, any player $i$ updates its strategy based on the payoff of a random target $j$ and itself, with a probability of $i$ imitating $j$ decided by the following Fermi function,
\begin{align}\label{eq:Fermi_total}
p_{ij} = \frac{1}{1+e^{\beta(\pi_i-\pi_j)}},
\end{align}
where $\beta$ is the selection strength. If $\beta$ approaches 0, the probability approaches $1/2$, meaning $i$ randomly chooses between retaining its strategy or imitating $j$'s strategy. If $\beta$ approaches infinity, the probability approaches $1$ if $j$'s payoff is higher than $i$'s, meaning $i$ will always imitate $j$'s strategy. Besides, players may mutate to take a random strategy with a probability of $\mu \in [0,1]$.

Since this paper considers several different strategy settings, the update rule varies accordingly while using similar functions to \eqref{eq:Fermi_total}. The specific nuance will be described in the following sections along with the introduction of different strategy settings.

\subsection{Binary Strategy} 
Firstly, we consider the most basic binary strategy setting, where there exist only two types of players, Cooperator ($C$)  and Defector ($D$). 
In this case, the strategy of player $i \in N$ in a game $g_V$ with $V \in \mcal{V}^k_i, k\in \{p,l,g\}$ is $s_i^V=\frac{o_i^k}{|\mcal{V}^k_i|}$ if $i$ cooperates, and 0 otherwise.
In other words, any player $i$ decides whether it cooperates or defects for all games it participates in. The endowment of the cooperators is then split evenly as contribution for games at each level it participates in. 
Moreover, instead of contributing all endowment, it is also possible to consider the more general case where the cooperators only use a fixed proportion of its endowment as contribution, which has been proved to be conducive to cooperation in some cases \citep{Zhao2023}.  This leads to the following strategy for any player $i$ in a game $g_V$ with $V \in \mcal{V}^k_i, k\in \{p,l,g\}$:
\begin{align}\label{eq:str_bin}
s_i^V=\left\{\begin{array}{ll}
\frac{\sigma o_i^k}{|\mcal{V}^k_i|} & \text{if $i$ cooperates,}  \\
0 & \text{otherwise,}
\end{array}
\right.
\end{align}
where $\sigma \in (0,1]$ is the overall percentage of endowment any player is willing to contribute. 

In this case, on a periodic square lattice, the payoff of a player depends on its own strategy, the number of cooperators among its first-order neighbors (for pairwise games) and second-order neighbors (for local games), and the whole population (for the global game). The payoff can then be expressed as a function of number of cooperators in each scope. If the number of cooperators in a set $V$ is denoted by $n_C^V$, and according to \eqref{eq:po} - \eqref{eq:po_g} the payoff of a cooperator $i$ at each level is then 
\begin{align}
\pi^p_i &=\sum\limits_{V \in \mcal{V}^p_i}\frac{r^pn_C^{V}\sigma o_i^p}{|V|}=\frac{r^p}{2}(|N(i)|+n_C^{N(i)}-2)\sigma o_i^p\\
\pi^l_i &= \sum\limits_{V \in \mcal{V}^l_i}\frac{r^ln_C^{V}\sigma o_i^l}{|V|}\\
\pi^g_i &=\frac{r^gn_C^{N}\sigma}{3|N|}.
\end{align}
the payoff of a defector $i$ at each level is then 
\begin{align}
\pi^p_i &=\frac{1}{3}+\sum\limits_{V \in \mcal{V}^p_i}\frac{r^pn_C^{V}\sigma o_i^p}{|V|}=\frac{1}{3}+\frac{r^p}{2}n_C^{N(i)}\sigma o_i^p\\
\pi^l_i &= \frac{1}{3}+\sum\limits_{V \in \mcal{V}^l_i}\frac{r^ln_C^{V}\sigma o_i^l}{|V|}\\
\pi^g_i &=\frac{1}{3}+\frac{r^gn_C^{N}\sigma}{3|N|}.
\end{align}

The update rule for binary strategy is that a player $i$ simply imitates $j$ according to the probability decided by the Fermi function \eqref{eq:Fermi_total}.

From the numerical results, it can be observed that at the boundary of clusters of cooperators and defectors, the imitation probability of the focus player does not change with the results of the global game. The reason is that even if the profit rate $r_g$ or the total number of cooperators change, it essentially inflicts the same payoff difference on all players at the boundary as long as they do not change themselves, and thus the payoff difference between the focus player and the target player does not change.
\subsection{Level-Based Strategy}
Secondly, in level-based strategy setting, it is assumed that players may employ the same strategy within games of the same level, while having the flexibility to vary their strategies from one level to another. In this case, there are 8 types of players, since player can choose from two strategies for each level. We use $CCC$, $CCD$, $CDC$, $CDD$, $DCC$, $DCD$, $DDC$, $DDD$ to denote the strategies for different types, each of which is composed of strategies (C or D) in the order of pairwise, local and global games, respectively. For instance, $CCD$ represents the type of players cooperate in pairwise games and local games, and defect in the global game.

As player strategies are level-based in this case, the strategy of any player $i$ in a game $g_V$ with $V \in \mcal{V}^k_i, k\in \{p,l,g\}$ can be represented by \eqref{eq:str_bin} with a small modification:
\begin{align}\label{eq:str_bin_2}
s_i^V=\left\{\begin{array}{ll}
\frac{\sigma o^k_i}{|\mcal{V}^k_i|} & \text{if $i$ cooperates in }\mcal{V}^k_i,  \\
0 & \text{otherwise.}
\end{array}
\right.
\end{align}

Similar to the binary strategy case, players update their strategies by imitating a random target's strategies with a probability decided by \eqref{eq:Fermi_total}.
\section{Results}\label{sec:res}
\subsection{Experimental Setting} \label{sec:exp}
Computational simulations are used to explore the dynamics of the aforementioned multi-level game, with a focus on the stable states and phase transitions between stable strategy configurations. 
Experiments are conducted on populations of sizes 100 - 900. The population structure is a periodic 2-dimensional lattice, where each player has 4 neighbors. In each round, all players update their strategies simultaneously. Some parameters and their default values are described in Table \ref{tab:para}, for which if the value of a parameter is not specified, it will take the default value. 
Under each parameter setting, the game is played until the population reaches a stable state (up to $10^6$ rounds depending on the setting), and is repeated for 20 times on average. 
\begin{table*}
\caption{Parameters}\label{tab:para}
\begin{tabular}{cccl}
\hline
Parameter &  Range & Default Value& Description\\\hline
$r^p$ &$[1,+\infty]$ &$1.6$ &profit rate for pairwise games\\
$r^l$ &$[1,+\infty]$ &$4$ &profit rate for local games\\
$r^g$ &$[1,+\infty]$ &$5$ &profit rate for the global game\\
$\beta$ &$[0.01,+\infty]$ &$0.5$ &selection strength\\
\hline
\end{tabular}
\end{table*}

\subsection{Results with Binary Strategy}
\begin{figure}
\centering
\includegraphics[width=0.3\textwidth]{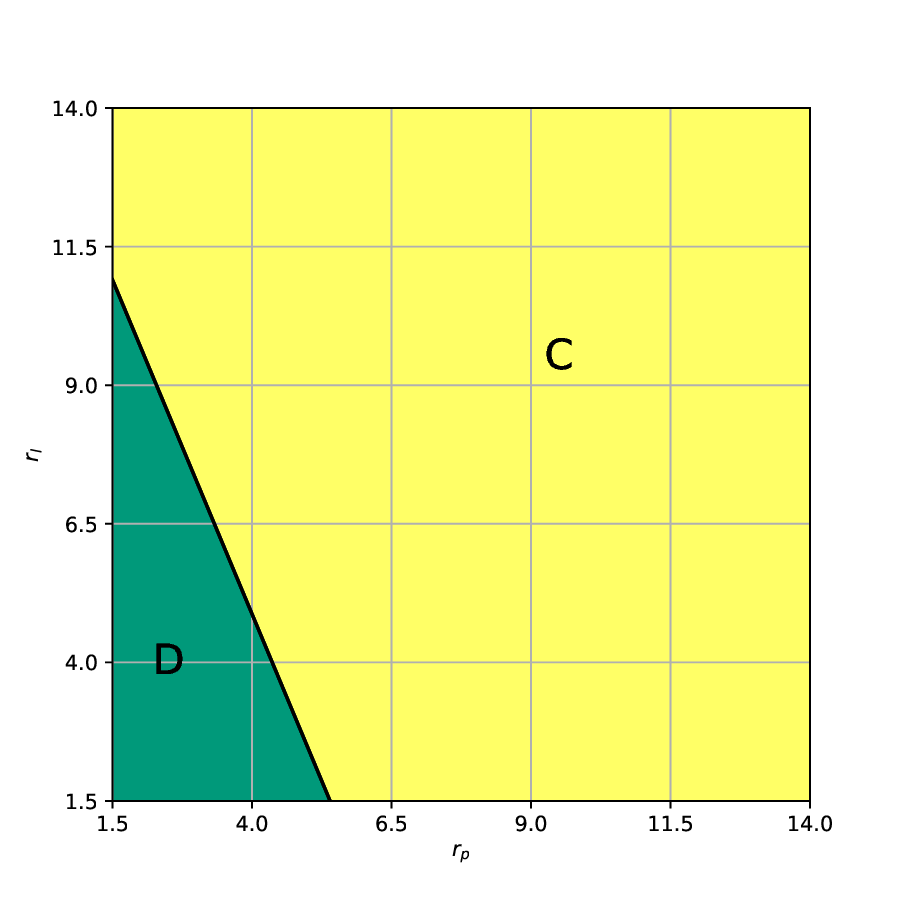}
\caption{Phase diagram of $r_p$ and $r_l$ under binary strategy.} \label{fig:phase_diagram_rg5}
\end{figure}

\begin{figure}
\centering
\includegraphics[width=0.5\textwidth]{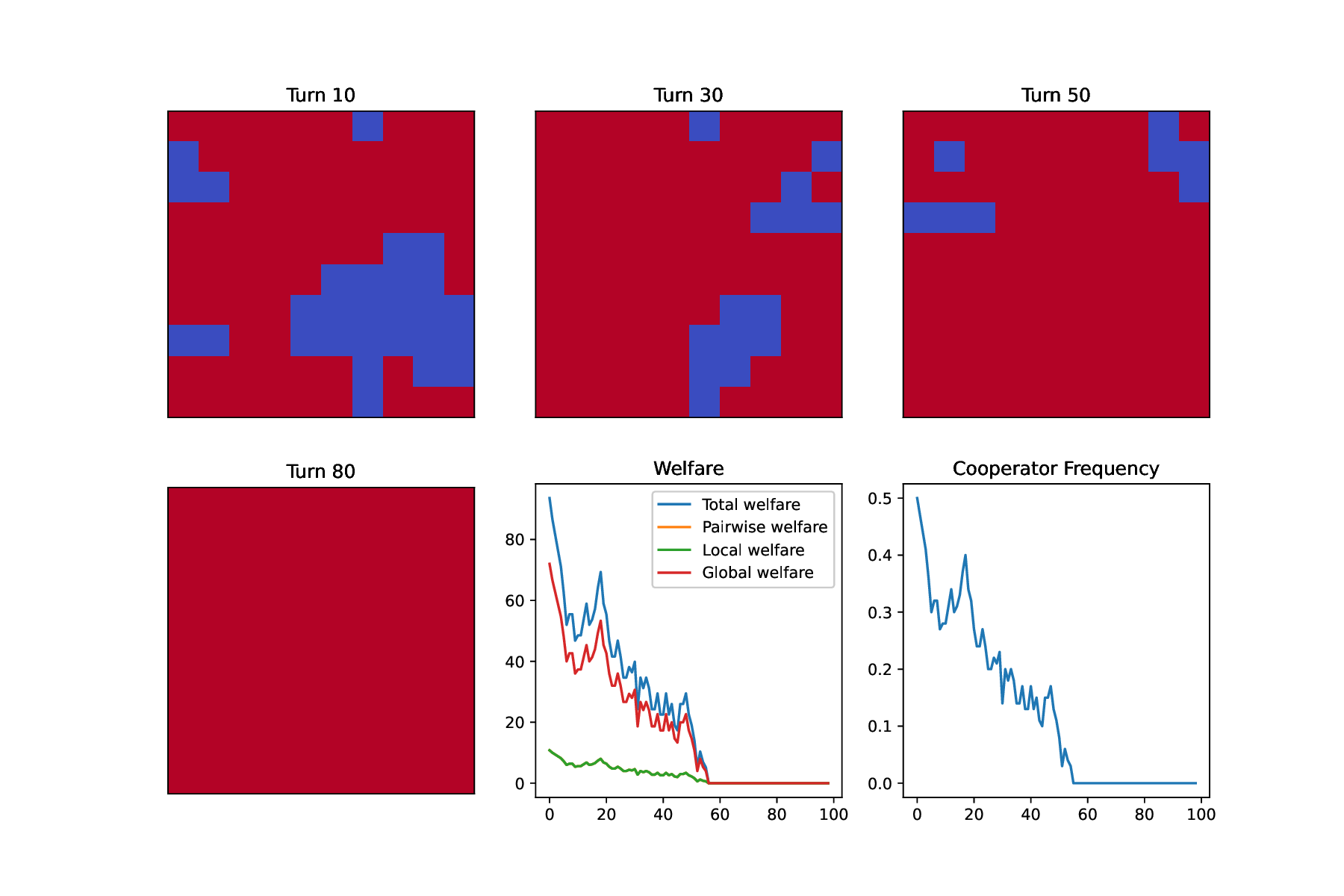}
\caption{Population evolving into D phase with $r_p=1.6, r_l=4$.} \label{fig:bin_D_4_16}
\end{figure}

\begin{figure}
\centering
\includegraphics[width=0.5\textwidth]{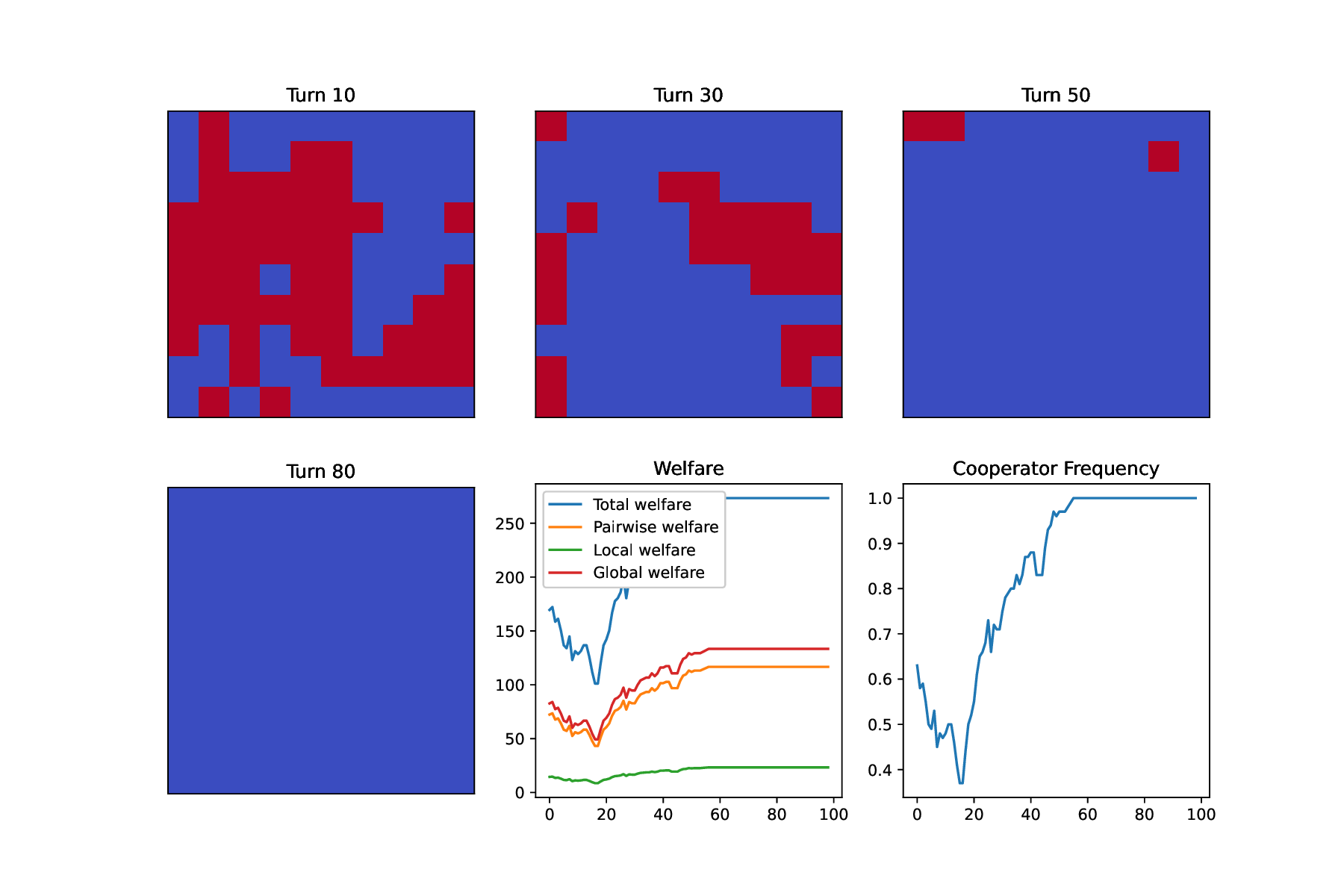}
\caption{Population evolving into C phase with $r_p=4.5, r_l=4.5$.} \label{fig:bin_C_45_45}
\end{figure}

\begin{figure}
\centering
\includegraphics[width=0.5\textwidth]{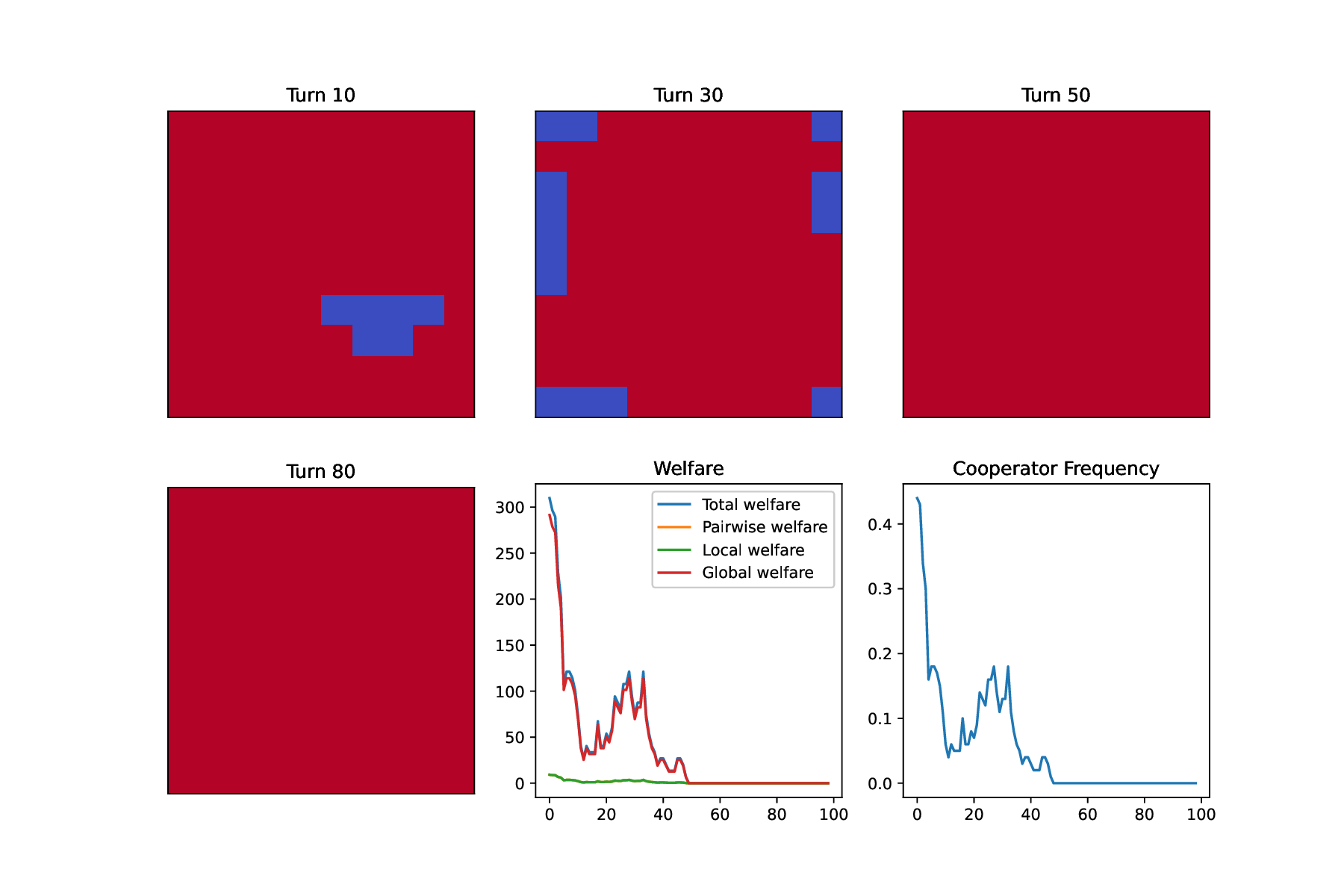}
\caption{Population evolving into D phase with $r_p=1.6, r_l=4, r_g=20$.} \label{fig:bin_D_4_16_rg20}
\end{figure}

Under the setting of binary strategy, experimental results show that the phase transition of cooperation level in the population is relatively intuitive, that is positively related to the change of profit rates at different levels. The increase of the profit rates of pairwise games and locals games eventually leads to a phase transition in the stable states of evolutionary dynamics from defection to cooperation at a certain tipping point, as shown in Figure \ref{fig:phase_diagram_rg5}. Figures \ref{fig:bin_D_4_16} and \ref{fig:bin_C_45_45} demonstrate the evolutionary processes of two instances under the settings of $r_p=1.6, r_l=4$ and $r_p=4.5, r_l=4.5$, respectively. 

Furthermore, it is particularly interesting that the profit rate of the global game has little to no impact on whether cooperators or defectors take over the population in our simulation results, as depicted in Figures \ref{fig:bin_D_4_16_rg20} (comparing to Figures \ref{fig:bin_D_4_16}). 

In order to reveal the mechanism that leads to this phenomenon, we delved into the details of the evolutionary process and perform some numerical simulation on the boundary of cooperators and defectors to see how the payoff and imitation probability changes in different settings. We find that the increase in global profit rate does not sway players' cooperation tendencies as expected, since their decisions are focused on localized regions. 

In general, cooperators gain an evolutionary advantage over defectors and are more likely to be imitated by neighboring defectors only when the profits from cooperators in the games are able to compensate their contribution and result in a higher payoff compared to the defectors in their neighbors. 

\begin{figure}
\centering
\begin{subfigure}{0.11\textwidth}
\centering
\includegraphics[width=\textwidth]{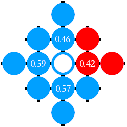}
\caption{}\label{fig:1645example_case1}
\end{subfigure}
\begin{subfigure}{0.11\textwidth}
\centering
\includegraphics[width=\textwidth]{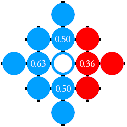}
\caption{}\label{fig:1645example_case2}
\end{subfigure}
\begin{subfigure}{0.11\textwidth}
\centering
\includegraphics[width=\textwidth]{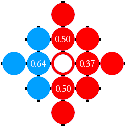}
\caption{}
\end{subfigure}
\begin{subfigure}{0.11\textwidth}
\centering
\includegraphics[width=\textwidth]{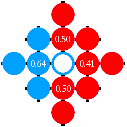}
\caption{} \label{fig:1645example_case4}
\end{subfigure}
\begin{subfigure}{0.11\textwidth}
\centering
\includegraphics[width=\textwidth]{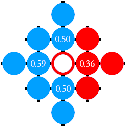}
\caption{}\label{fig:1645example_case5}
\end{subfigure}
\begin{subfigure}{0.11\textwidth}
\centering
\includegraphics[width=\textwidth]{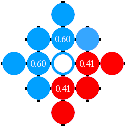}
\caption{}
\end{subfigure}
\begin{subfigure}{0.11\textwidth}
\centering
\includegraphics[width=\textwidth]{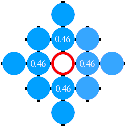}
\caption{}
\end{subfigure}
\begin{subfigure}{0.11\textwidth}
\centering
\includegraphics[width=\textwidth]{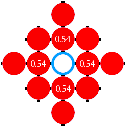}
\caption{}\label{fig:1645example_case8}
\end{subfigure}
\begin{subfigure}{0.5\textwidth}
\centering
\includegraphics[width=0.7\textwidth]{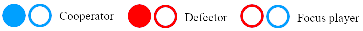}
\end{subfigure}
\caption{Example cases with $r_p = 1.6, r_l=4, r_g=5$ under binary strategy. The figures depict a local region containing a focus player and its first-order and second-order neighbors. 
 The numbers on the neighbors of the focus player are the probabilities that the focus player imitates them if they are chosen as the target. For comparison, the imitation probabilities of all neighbors are presented regardless of its strategy, while in the actual game imitating a player of the same strategy will not change the game state.}\label{fig:1645example}
\end{figure}

\begin{figure}
\centering
\begin{subfigure}{0.11\textwidth}
\centering
\includegraphics[width=\textwidth]{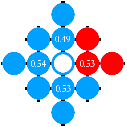}
\caption{}
\end{subfigure}
\begin{subfigure}{0.11\textwidth}
\centering
\includegraphics[width=\textwidth]{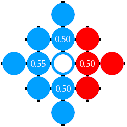}
\caption{}
\end{subfigure}
\begin{subfigure}{0.11\textwidth}
\centering
\includegraphics[width=\textwidth]{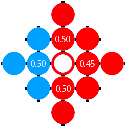}
\caption{}
\end{subfigure}
\begin{subfigure}{0.11\textwidth}
\centering
\includegraphics[width=\textwidth]{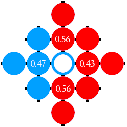}
\caption{} \label{fig:15155example_case4}
\end{subfigure}
\begin{subfigure}{0.11\textwidth}
\centering
\includegraphics[width=\textwidth]{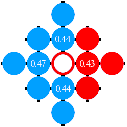}
\caption{}\label{fig:15155example_case5}
\end{subfigure}
\begin{subfigure}{0.11\textwidth}
\centering
\includegraphics[width=\textwidth]{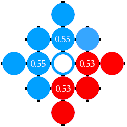}
\caption{}
\end{subfigure}
\begin{subfigure}{0.11\textwidth}
\centering
\includegraphics[width=\textwidth]{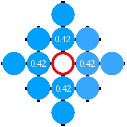}
\caption{}
\end{subfigure}
\begin{subfigure}{0.11\textwidth}
\centering
\includegraphics[width=\textwidth]{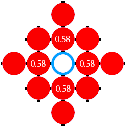}
\caption{}\label{fig:15155example_case8}
\end{subfigure}
\begin{subfigure}{0.5\textwidth}
\centering
\includegraphics[width=0.7\textwidth]{num_legend_new.eps}
\end{subfigure}
\caption{Example cases with $r_p = 1.5, r_l=1.5, r_g=5$ under binary strategy. The figures depict a local region containing a focus player and its first-order and second-order neighbors. The number on each neighbor of the focus player represents the probability that the focus player will imitate that neighbor if selected as the target. For comparison, the imitation probabilities of all neighbors are presented regardless of its strategy, while in the actual game imitating a player of the same strategy will not change the game state.}\label{fig:15155example}
\end{figure}

\begin{figure}
\centering
\begin{subfigure}{0.11\textwidth}
\centering
\includegraphics[width=\textwidth]{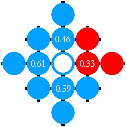}
\caption{}
\end{subfigure}
\begin{subfigure}{0.11\textwidth}
\centering
\includegraphics[width=\textwidth]{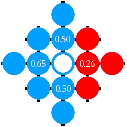}
\caption{}
\end{subfigure}
\begin{subfigure}{0.11\textwidth}
\centering
\includegraphics[width=\textwidth]{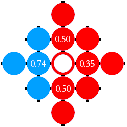}
\caption{}
\end{subfigure}
\begin{subfigure}{0.11\textwidth}
\centering
\includegraphics[width=\textwidth]{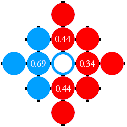}
\caption{}
\end{subfigure}

\begin{subfigure}{0.11\textwidth}
\centering
\includegraphics[width=\textwidth]{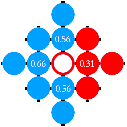}
\caption{}
\end{subfigure}
\begin{subfigure}{0.11\textwidth}
\centering
\includegraphics[width=\textwidth]{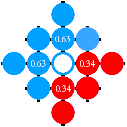}
\caption{}
\end{subfigure}
\begin{subfigure}{0.11\textwidth}
\centering
\includegraphics[width=\textwidth]{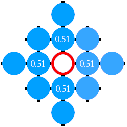}
\caption{}\label{fig:45455example_case7}
\end{subfigure}
\begin{subfigure}{0.11\textwidth}
\centering
\includegraphics[width=\textwidth]{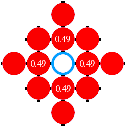}
\caption{}
\end{subfigure}
\begin{subfigure}{0.5\textwidth}
\centering
\includegraphics[width=0.7\textwidth]{num_legend_new.eps}
\end{subfigure}
\caption{Example cases with $r_p = 4.5, r_l=4.5, r_g=5$ under binary strategy. The figures depict a local region containing a focus player and its first-order and second-order neighbors. 
 The number on each neighbor of the focus player represents the probability that the focus player will imitate that neighbor if selected as the target. For comparison, the imitation probabilities of all neighbors are presented regardless of its strategy, while in the actual game imitating a player of the same strategy will not change the game state.} \label{fig:45455example}
\end{figure}

\begin{figure}
\centering
\begin{subfigure}{0.11\textwidth}
\centering
\includegraphics[width=\textwidth]{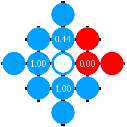}
\caption{}
\end{subfigure}
\begin{subfigure}{0.11\textwidth}
\centering
\includegraphics[width=\textwidth]{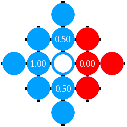}
\caption{}
\end{subfigure}
\begin{subfigure}{0.11\textwidth}
\centering
\includegraphics[width=\textwidth]{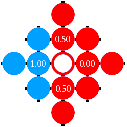}
\caption{}
\end{subfigure}
\begin{subfigure}{0.11\textwidth}
\centering
\includegraphics[width=\textwidth]{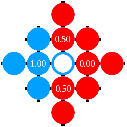}
\caption{} \label{fig:1645ss10example_case4}
\end{subfigure}
\begin{subfigure}{0.11\textwidth}
\centering
\includegraphics[width=\textwidth]{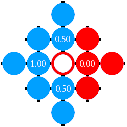}
\caption{}\label{fig:1645ss10example_case5}
\end{subfigure}
\begin{subfigure}{0.11\textwidth}
\centering
\includegraphics[width=\textwidth]{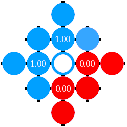}
\caption{}
\end{subfigure}
\begin{subfigure}{0.11\textwidth}
\centering
\includegraphics[width=\textwidth]{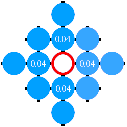}
\caption{}
\end{subfigure}
\begin{subfigure}{0.11\textwidth}
\centering
\includegraphics[width=\textwidth]{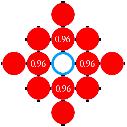}
\caption{}\label{fig:1645ss10example_case8}
\end{subfigure}
\begin{subfigure}{0.5\textwidth}
\centering
\includegraphics[width=0.7\textwidth]{num_legend_new.eps}
\end{subfigure}
\caption{Example cases with $r_p = 1.6, r_l=4, r_g=5$ and $\beta = 10$ under binary strategy. The figures depict a local region containing a focus player and its first-order and second-order neighbors. 
 The numbers on the neighbors of the focus player are the probabilities that the focus player imitates them if they are chosen as the target. For comparison, the imitation probabilities of all neighbors are presented regardless of its strategy, while in the actual game imitating a player of the same strategy will not change the game state.}\label{fig:1645ss10example}
\end{figure}

We report a set of quintessential results to reflect the general trends. Particularly, Figures \ref{fig:1645example} - \ref{fig:1645ss10example} show imitation probabilities of a focus player in different neighbor environments under different settings. Figure \ref{fig:1645example} depicts the results under the setting of $r_p =1.6, r_l = 4$ and $r_g = 5$; Figure \ref{fig:15155example} depicts the results under the setting of $r_p =1.5, r_l = 1.5$ and $r_g = 5$;  Figure \ref{fig:45455example} depicts the results under the setting of $r_p =4.5, r_l = 4.5$ and $r_g = 5$; Figure \ref{fig:1645ss10example} depicts the results under the setting of $r_p =1.6, r_l = 4$, $r_g = 5$ and a higher selection strength of $\beta =10$. 

Note that some assumptions are necessary to present the results. For instance, the number of cooperators in the local games of the secondary neighbors of the focus players are assigned according to the boundaries. In addition, the number of cooperators in the global game is assumed to be $50\%$ of the total population, which will also be proved to be generally indifferent to the results mathematically.

According to the results, it can be observed that when the profit rates are relatively low (in Figure \ref{fig:15155example}), cooperators have a high probability ($>50\%$) to imitate defectors in its neighbors (e.g., Figure \ref{fig:15155example_case4}), especially when they have cooperative neighbors. Defectors have a lower probability ($<50\%$)  to imitate cooperators in its neighbors  (e.g., Figure \ref{fig:15155example_case5}), especially when the focus player is surrounded by defectors in Figure \ref{fig:15155example_case8}. 

When the profit rates increase, cooperators will gain advantage in the imitation process, which is reflected by having a higher probability of being imitated by defectors and a lower probability of imitating defectors (see differences of the same cases in Figures \ref{fig:1645example}, \ref{fig:15155example}, and \ref{fig:45455example}).
This selection effect is even more prominent with a higher selection strength, as shown in Figure \ref{fig:1645ss10example} comparing to Figure \ref{fig:1645example}. 

From the numerical analysis, some general patterns that are invariant throughout different parameter settings can be observed and summarized. 

Firstly, payoff of the global game does not affect the imitation probability in such local analysis. The reason is that under our experimental setting, the change in the number of total cooperators affects all players equally, regardless of their position and strategy. Consequently, as the imitation probability depends only on the payoff difference of the focus player and the target player according to \eqref{eq:Fermi_total}, it remains the same as the payoff difference remains unchanged. Therefore, even if the profit rate of the global game $r_g$ and the number of cooperators in the population changes drastically, the imitation probability within a local scope will remain unchanged as long as the players within the local scheme do not alter their strategies. As a result, different values of $r_g$ as well as number of cooperators in the whole population lead to the same numerical results for all cases, which thus are omitted.

Secondly, an intuitive pattern is that a player tends to have a higher payoff when there are more cooperators in its neighbors, and thus tends to have a higher probability to be imitated if it is the target, and have a lower probability to imitate others if it is the focus player (e.g., see Figures \ref{fig:1645example_case1} and \ref{fig:1645example_case2}). This pattern can also be enhanced by higher profit rates (see Figure \ref{fig:45455example}), as well as a higher selection strength (see Figure \ref{fig:1645ss10example}). The reason behind this pattern is the nature of public goods games played at all levels, where cooperators benefit (or contribute to the payoff of) all players within the game. However, this pattern may have some minor exceptions since a player's payoff also depends on its secondary neighbors to a lesser extent.

In cases such as Figure \ref{fig:45455example_case7}, the focus player has a higher probability ($>50\%$) to imitate cooperators in its neighbors, even though it has more cooperative neighbors than all its neighbors. The reason is that, its neighbor cooperators benefit more from the local games and other pairwise games they engage in, while any game the focus player participates automatically has a defector, i.e., itself. Therefore, the focus player has a lower payoff than its cooperative neighbors overall. In addition, since the parameter setting ($r_p=4.5, r_l =4.5$) locates near the boundary of the phase diagram depicted in Figure \ref{fig:phase_diagram_rg5}, the imitation probability is slightly above $50\%$. This is also reflected in the evolutionary process, as shown in Figure \ref{fig:bin_C_45_45}, where the cooperator frequency experienced a dip (due to the instability of imitation at a probability close to $50\%$) before eventually rising to 1.

From an evolutionary point of view, these patterns lead to the increase of the fitness of cooperators against defectors at the boundary cases when the pairwise and local profit rates increase. Therefore, the phase diagrams in Figure \ref{fig:phase_diagram_rg5} show that the evolutionary results of the population shifts from D-phase to C-phase when the pairwise and local profit rates are sufficiently large.

It is also worth noting that the cases where the population evolves to an all-cooperator state have quite unrealistic $r_p$ and $r_l$ values, i.e., the values of profit rates are larger than the number of participants in the games. It shows to an extent that cooperation is difficult to achieve and maintain in such scenarios. 
\subsection{Results with Level-Based Strategy}
\begin{figure}
\centering
\includegraphics[width=0.3\textwidth]{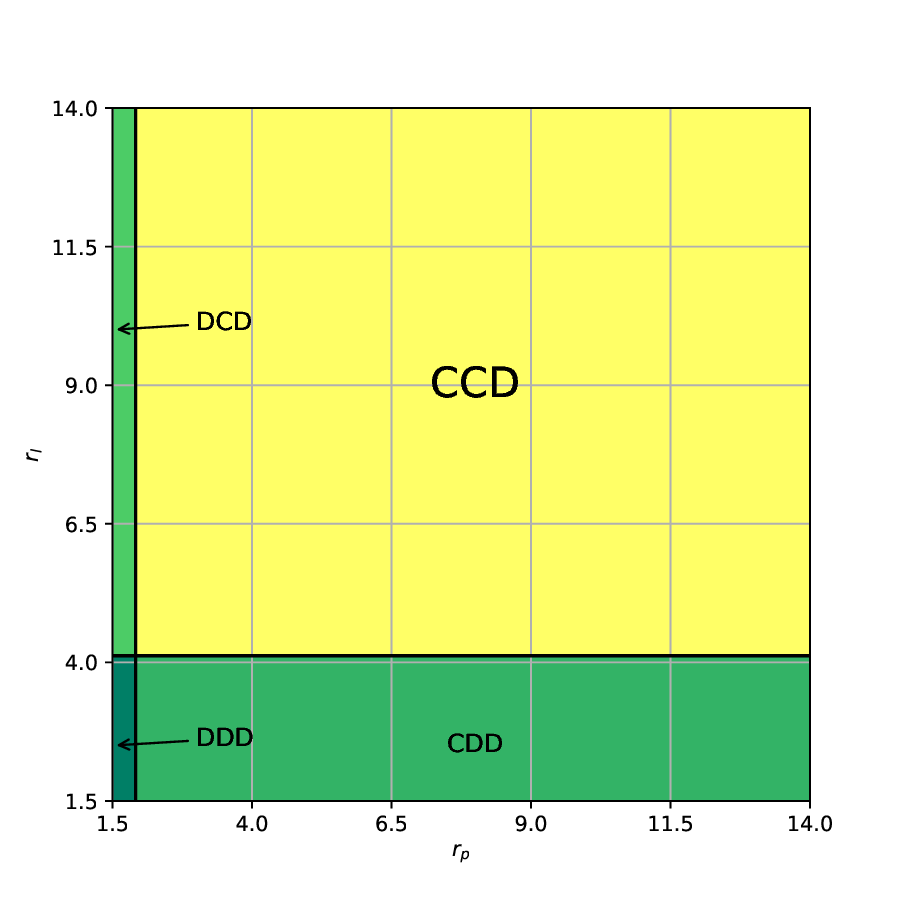}
\caption{Phase diagram of $r_p$ and $r_l$ with $r_g = 5$ and $\beta =100$}\label{fig:phase_levelbased_100_rg5}
\end{figure}

With level-based strategy, we perform computational simulations under similar settings to the previous case. As shown by the phase diagram in Figure \ref{fig:phase_levelbased_100_rg5},  
it also has a general trend that cooperation is promoted by higher values of $r_p$ and $r_l$. This pattern can also be reinforced with a higher selection strength $\beta$ to reduce the randomness of the imitation. Particularly, the population tends to evolve towards a one-strategy population, and under different settings of $r_p$ and $r_l$, different types of players will take over the population. Moreover, the value of $r_p$ decides whether the population will be taken over by a strategy that cooperates in pairwise games, which results in a phase transition from DCD to DCC or from DDD to DDC in the phase diagram.  The value of $r_l$ decides whether the population will be taken over by a strategy that cooperates in local games, which results in a phase transition from DDD to DCD or from DDC to DCC in the phase diagram. Once again, the observed pattern is consistent over different values of $r_g$. 

Comparing to the binary strategy case, players have the option to change the strategy of a level independently from strategies of other two levels, and thus are able to adjust their strategies of each level separately according to their payoffs. 

\begin{figure}
\centering
\includegraphics[width=0.5\textwidth]{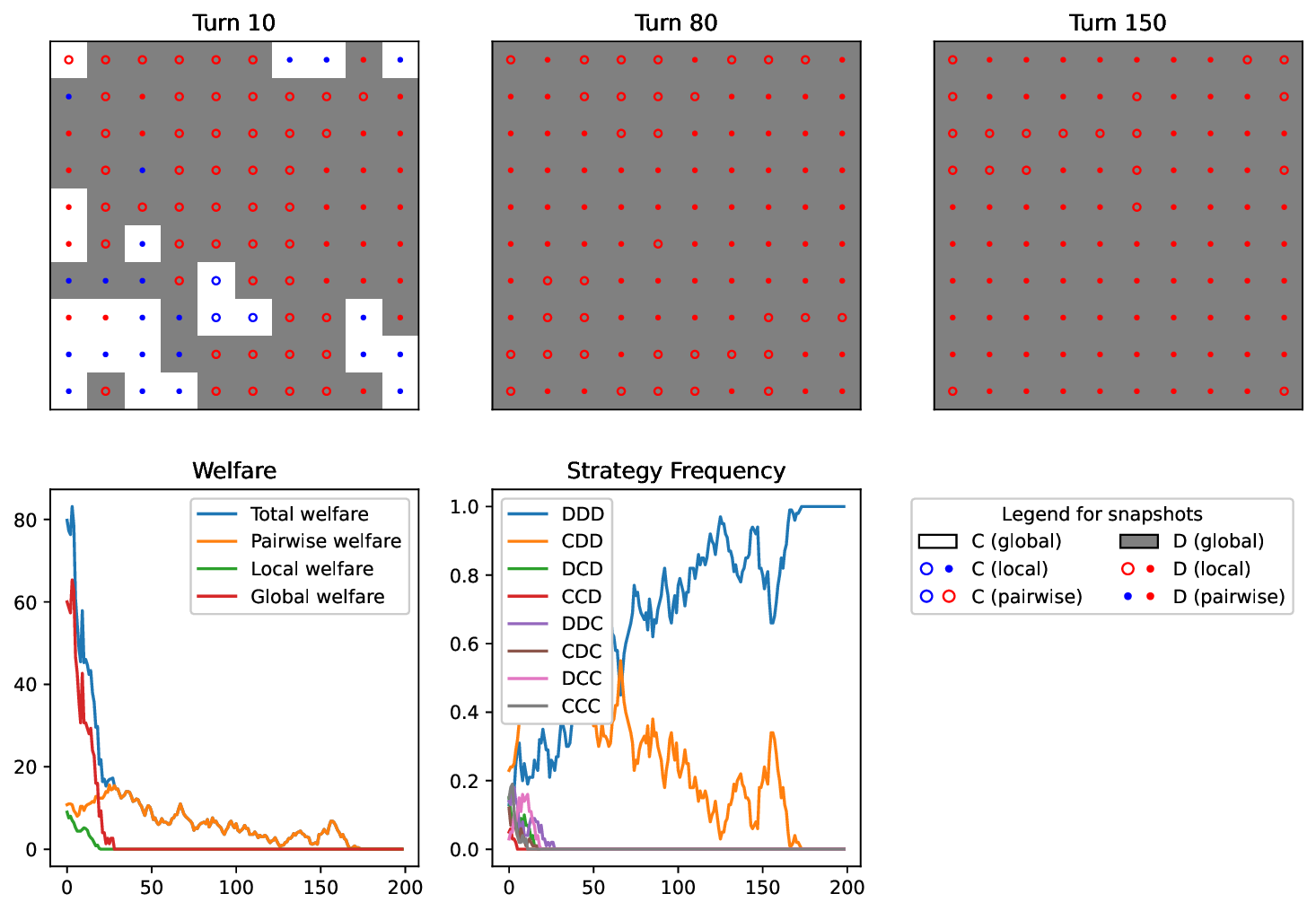}
\caption{Population evolving into DDD phase with $r_p=1.6, r_l=4,  r_g=5, n=100$.}\label{fig:snapshot_level_100_4_16}
\end{figure}

\begin{figure}
\centering
\includegraphics[width=0.5\textwidth]{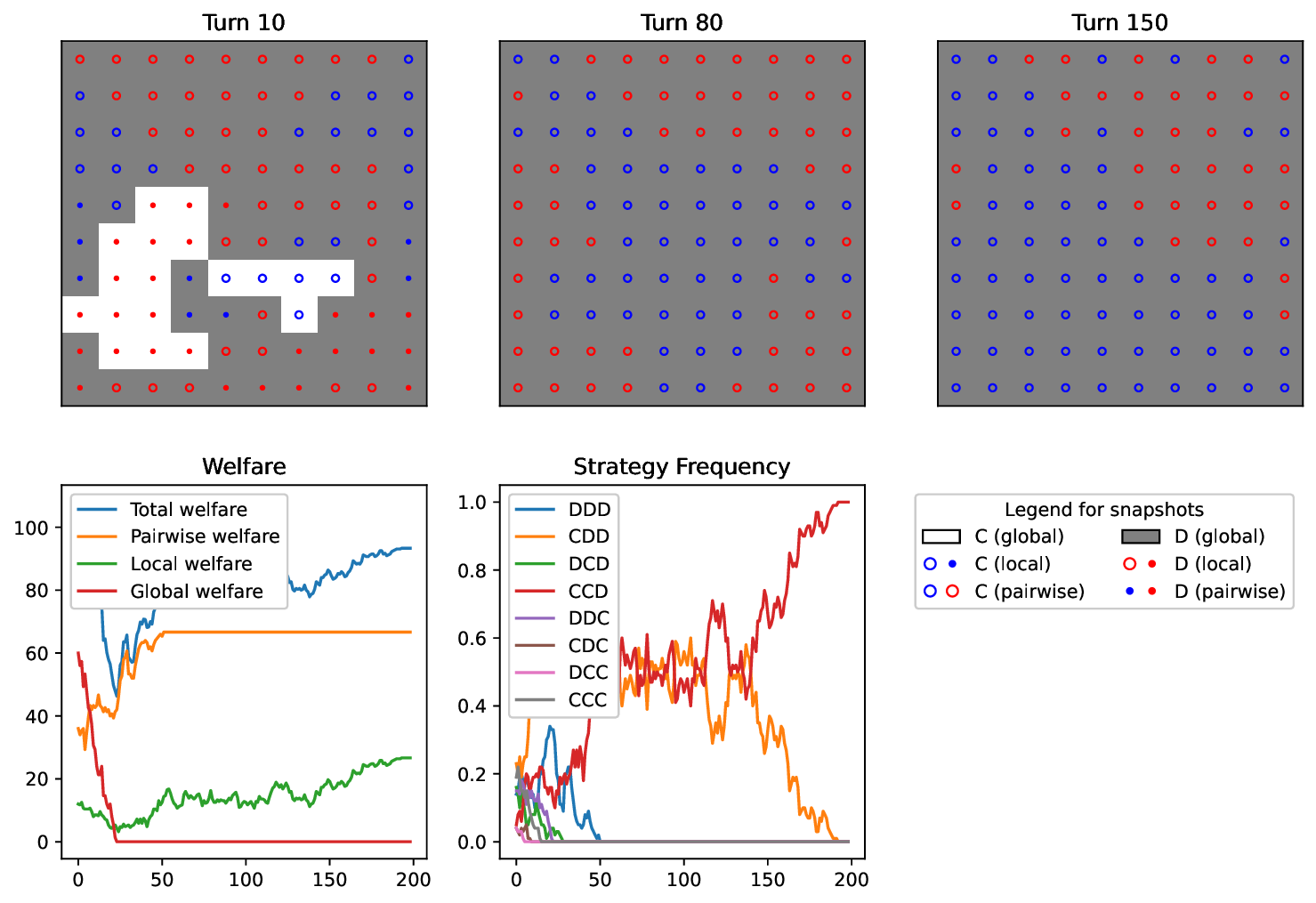}
\caption{Population evolving into CCD phase with $r_p=3, r_l=5,  r_g=5, n=100$.}\label{fig:snapshot_level_100_5_3}
\end{figure}

\begin{figure}
\centering
\includegraphics[width=0.5\textwidth]{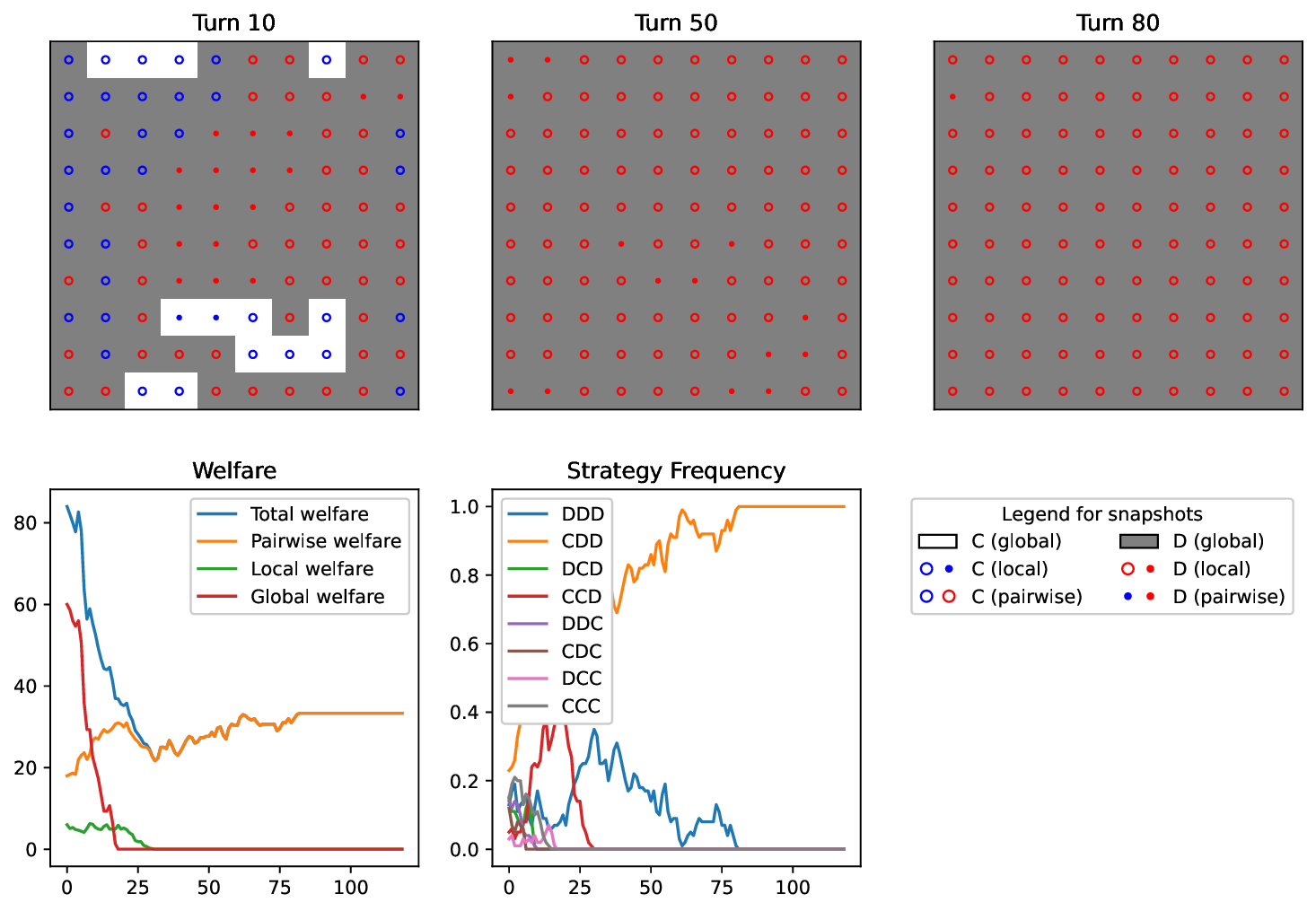}
\caption{Population evolving into CDD phase with $r_p=2, r_l=3,  r_g=5, n=100$.}\label{fig:snapshot_level_100_3_2}
\end{figure}

\begin{figure}
\centering
\includegraphics[width=0.5\textwidth]{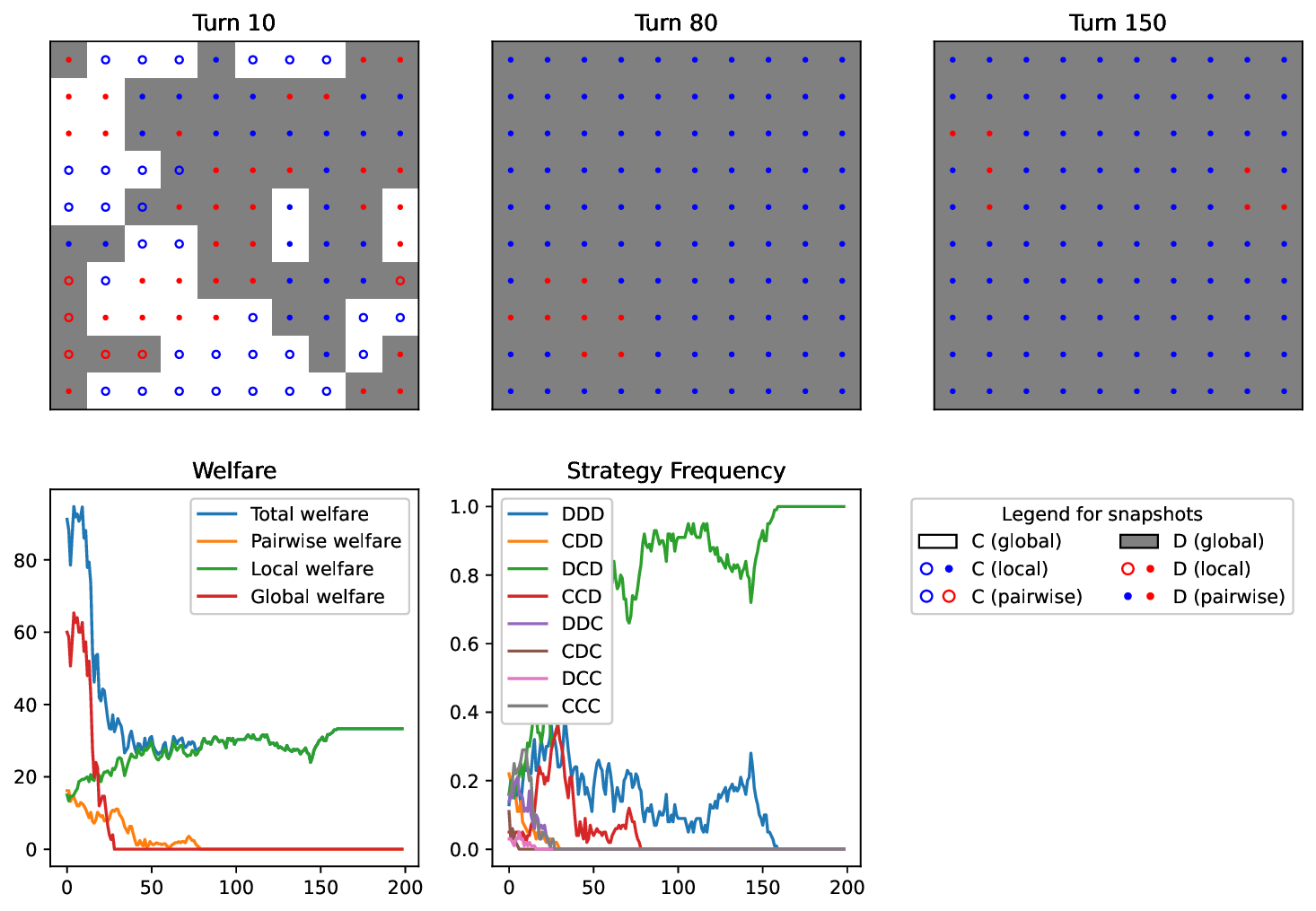}
\caption{Population evolving into DCD phase with $r_p=1.9, r_l=6,  r_g=5, n=100$.}\label{fig:snapshot_level_100_6_19}
\end{figure}

One notable aspect of the simulation results is that only the type of players which defect in the global game can take over the population. As demonstrated in Figures \ref{fig:snapshot_level_100_4_16}-\ref{fig:snapshot_level_100_3_2}, strategies that cooperate in the global game are eliminated very early in the evolutionary process, regardless of the settings. In most cases, the population will first evolve to a two-strategy competition state, and the winning strategy of the two will take over the population eventually. Particularly, Figure \ref{fig:snapshot_level_100_4_16} corresponds to the parameter setting located in the DDD phase in Figure \ref{fig:phase_levelbased_100_rg5}, while being close to the boundaries between all four phases. For this case, we may observe a two-strategy competition during the evolutionary process between DDD and any of the other three strategies. Figure \ref{fig:snapshot_level_100_5_3} corresponds to the parameter setting located in the CCD phase in Figure \ref{fig:phase_levelbased_100_rg5}, which is close to the boundary between CCD and CDD, and hence we observe these two strategies competing during the evolutionary process. For similar reasons, Figure \ref{fig:snapshot_level_100_5_3} shows a two-strategy competition between CDD and DDD before evolving to an all-CDD state. Figure \ref{fig:snapshot_level_100_6_19} depicts a case of minor exception we encounter, where two-strategy competition is between DCD and DDD, instead of DCD and CCD in most cases. Additionally, it can also be observed that, initially the population evolves to a three-strategy competition among DCD, CCD and DDD, before it shifts to the two-strategy competition. 
 
Further experiments show that the obtained results are consistent over different values of $r_g$. The reason of this phenomenon could be that cooperating in the global game would only yield disadvantage at the boundaries of player clusters. 
This to some extent is consistent with the results under binary strategy setting, as the global game has little impact on either evolutionary result or the phase shift of the population.

To summarize, the experimental results with both binary strategy and level-based strategy show intuitive patterns regarding local and pairwise games, where the population shifts to a state dominated by cooperation when the corresponding profit rate increases to a tipping point. However, the global game profit hardly has any impact on the population state or individual strategy choice as shown in simulation results and numerical results. 
\section{Conclusion}\label{sec:conclusion}
This paper proposes a multi-level game model to analyze  complex interactions among players in multi-lateral scenarios such as nations facing climate change.  The model incorporates three levels: pairwise, local and global games are considered, with all levels represented as n-player public good games. Two types of player strategies are studied: binary strategy, where players choose the same strategy at all levels, and level-based strategy, where players can select different strategies for each level.

Through a combination of computational simulations and numerical analysis on populations with a periodic lattice structure, we obtained results showing both intuitive and counter-intuitive trends. On the one hand, increasing the profit rates of pairwise games and local games generally drives a phase shift in the population from defection to cooperation, in both binary strategy and level-based strategy cases. Nonetheless, in the level-based strategy case, changes in profit rate more precisely affect cooperation at the corresponding level (i.e. changes in pairwise-game profit rate affect cooperation in pairwise games, and changes in local-game profit rate affect cooperation in local games). On the other hand, changes in global-game profit rate do not affect the overall cooperation level in the population, regardless of strategy type. More specifically, with binary strategy, results show that global-game profit rate does not affect the imitation probability at the boundary of cooperators and defectors, which prevents cooperation from being achieved by increasing the global profit rate. In the level-based strategy case, population tends to converge only to states where players all defect in the global game, whereas the pairwise and local game profits decide whether the dominating strategy cooperates at these two levels respectively.

The obtained results could have a few practical implications. First of all, in game interactions with such multi-level nature and local decision-making, it is generally more beneficial to invest in promoting cooperation at lower levels or within smaller scopes (such as pairwise and local levels in our case), than at higher or broader levels (such as the global game). Alternatively, fostering collaboration through simultaneous agreements across multiple levels may strengthen global cooperation, especially when participants focus on localized considerations in their decision-making. Besides, cooperation especially in a large population is extremely challenging to maintain, which has also been pointed out in prior computational and experimental studies \citep{Milinski2008, Wang2020, Zhao2023}. Additionally, when the profit rate is realistic, it seems necessary to adopt mechanisms such as reward and sanction to effectively promote cooperation in a large population.

Although this work extends the study of interactions and decision-making in multi-level systems, it has several limitations. For instance, a broader range of game types, population structures, and player strategies could be explored within this multi-level framework. Additionally, effective approaches and mechanisms for achieving cooperation in such complex scenarios warrant further investigation. Moreover, the potential for modeling player decisions with a global perspective in these systems remains unexplored, and it also raises the question of whether such an approach could address the issue of global game indifference. Nevertheless, we hope this work inspires future research to analyze complex interactions using multi-level game models and offers valuable insights for addressing challenges such as global climate change.
\section{Acknowledgment}
This work is supported by National Natural Science Foundation of China, Grant (No. 72101189 and 72031009), the National Social Science Foundation of China (No. 20\&ZD058), and Zhengzhou City Special Project for Collaborative Innovation in Science and Technology (Soft Science, No. 21ZZXTCX26).
\bibliography{multi-level}
\end{document}